# Visualizing Field-free Deterministic Magnetic Switching of all–van der Waals Spin-Orbit Torque System Using Spin Ensembles in Hexagonal Boron Nitride

Xi Zhang[1,+], Jingcheng Zhou[1,+], Chaowei Hu[2], Kuangyin Deng[3], Chuangtang Wang[4], Nishkarsh Agarwal[5], Hanshang Jin[6], Faris A. Al-Matouq[1], Stelo Xu[6], Roshan S. Trivedi[1], Senlei Li[1], Sumedh Rathi[1], Hanyi Lu[7,1], Zhigang Jiang[1], Valentin Taufour[6], Robert Hovden[5,8], Liuyan Zhao[4], Ran Cheng[3,9], Xiaodong Xu[2,10], Jiun-Haw Chu[2], Chunhui Rita Du[1,*], and Hailong Wang[1,*]

[1]School of Physics, Georgia Institute of Technology, Atlanta, Georgia 30332, USA
[2]Department of Physics, University of Washington, Seattle, Washington 98105, USA
[3]Department of Electrical and Computer Engineering, University of California, Riverside, California 92521, USA
[4]Department of Physics, University of Michigan, Ann Arbor, Michigan 48109, USA
[5]Department of Materials Science and Engineering, University of Michigan, Ann Arbor, Michigan 48109, USA
[6]Department of Physics and Astronomy, University of California, Davis, California 95616, USA
[7]Department of Physics, University of California, San Diego, La Jolla, California 92093, USA
[8]Applied Physics Program, University of Michigan, Ann Arbor, Michigan 48109, USA
[9]Department of Physics and Astronomy, University of California, Riverside, California 92521, USA
[10]Department of Materials Science and Engineering, University of Washington, Seattle, Washington 98105, USA

[*]Corresponding authors: cdu71@gatech.edu, hwang3021@gatech.edu
[+]These authors contributed equally.

**Abstract**: Recently, optically active spin defects embedded in van der Waals (vdW) crystals have emerged as a transformative quantum sensing platform to explore cutting-edge materials science and quantum physics. Taking advantage of excellent solid-state integrability, this new class of spin defects can be arranged in controllable nanoscale proximity of target materials in vdW heterostructures, showing great promise for improving spatial resolution and field sensitivity of current sensing technologies. Building on this state-of-the-art measurement platform, here we report hexagonal boron nitride-based quantum imaging of field-free deterministic magnetic switching of room-temperature two-dimensional magnet $Fe_3GaTe_2$ in an all-vdW spin-orbit torque (SOT) system. By visualizing SOT-driven variations of nanoscale $Fe_3GaTe_2$ magnetic stray field profile under different conditions, we have revealed how the observed magnetic switching evolves from deterministic to indeterministic behavior due to the interplay between out-of-plane spins, in-plane spins and Joule heating. This understanding, which is otherwise difficult to access by conventional transport measurements, offers valuable insights on material design, testing, and evaluation of next-generation vdW spintronic devices.

Spin-orbit torques (SOTs) have been widely used for developing fast, scalable, and energy-efficient magnetization control in modern spintronic memory devices[1,2]. The recent flourishing van der Waals (vdW) crystals provide an appealing material playground to design, improve, and revolutionize the conventional SOT technologies[3–6]. Potential advantages associated with the emerging two-dimensional (2D) spintronic devices include highly engineerable interfacial conditions, tunable polarization of injected spin currents, convenient material co-integrations, and readily established magnetic vdW proximity, opening new pathways for building ultracompact spin logic circuits with tailored functionalities and hybrid performance[3,7]. The recently observed out-of-plane spin driven field-free deterministic magnetic switching of a perpendicular vdW magnet is a notable example in this catalog[8–11]. The elimination of auxiliary magnetic field(s) is particularly favorable for implementing scalable spin memory applications with reduced energy consumption[8,10–13].

Despite the remarkable progress made thus far, ongoing research on unconventional SOT-driven field-free magnetic switching in vdW heterostructures has mainly focused on magneto-transport measurements. These studies are susceptible to thermal, electromigration, and magnetoelastic artifacts[8–11], which can in principle be resolved by nanoscale imaging of magnetization switching details. However, implementation of such cutting-edge imaging technique(s) in all-vdW SOT systems remains an open challenge in the current state-of-the-art. This limitation hinders a comprehensive understanding of the fundamental magnetic switching mechanism and impedes future improvement of the device performance.

Here, we report hexagonal boron nitride (hBN)-based quantum imaging[14–17] of field-free deterministic magnetization switching in $WTe_2/Fe_3GaTe_2$/hBN vdW heterostructures. Out-of-plane polarized spin currents are generated by the spin source material $WTe_2$ with reduced crystallographic symmetry[8–10,18], which exerts unconventional SOTs on proximal perpendicular $Fe_3GaTe_2$ magnetization[10]. We show that robust deterministic magnetic switching can be realized in few-layer thick vdW magnet $Fe_3GaTe_2$ without the assistance of an external magnetic field. Nanoscale magnetic imaging was achieved by utilizing optically active spin defects embedded in the constituent hBN encapsulation layer[17,19–22]. Using wide-field quantum magnetometry techniques[14,15], we directly visualize the microscopic $Fe_3GaTe_2$ stray field profile as a function of electrical switching currents, revealing systematic evolutions of magnetic domains in response to in-plane and/or out-of-plane polarized spin current injection, which is not accessible by conventional electrical SOT measurements. Our study enriches the current understanding of SOT-induced magnetization dynamics in vdW heterostructures, providing valuable information for future design of energy-efficient 2D spin memory devices. Considering the extensive usage of hBN in preparing vdW nano-electronics[23,24], we share the optimism that the demonstrated vdW quantum sensing platform could be extended readily to a broad family of 2D stacking systems, opening new avenues for investigating nanoscale electromagnetic behaviors in 2D quantum matter[24].

We first review the vdW heterostructures used in the current study for hBN quantum sensing and electrical transport measurements as illustrated in Fig. 1a. Exfoliated $WTe_2$, $Fe_3GaTe_2$, and hBN nanoflakes with desirable thicknesses and lateral dimensions are picked up in sequence to form $WTe_2/Fe_3GaTe_2$/hBN stacking devices, and then released onto $Si/SiO_2$ substrates with patterned platinum (Pt) electrodes. The thicknesses of $WTe_2$ and $Fe_3GaTe_2$ layers are ~10 nm and ~7 nm to balance the interfacial spin accumulation and electrical shunting effect (see Supplementary Information Note 1 for details). The hBN encapsulation layer is utilized to prevent sample degradation and serves as a 2D quantum sensing platform as discussed below. It is worth mentioning that Fe3GaTe2 is a recently discovered vdW magnet with a bulk Curie temperature exceeding 350 K[25–27]. Few-layer-thick Fe3GaTe2 shows robust magnetic order with a perpendicular magnetic anisotropy above room temperature (see Supplementary Information Note 2 for details), serving as an ideal material candidate to develop functional vdW spintronic devices for practical applications[10,25,26]. WTe2 is the spin source material whose broken crystal symmetry allows generation of out-of-plane polarized spin currents when electrical charge currents flow along its low-symmetry crystallographic axis (*a*-axis)[8–10,18]. Figure 1b shows a top-down overview optical microscopy image of a prepared WTe2/Fe3GaTe2/hBN vdW stacking device (device A), and Fig. 1c shows the cross-sectional high angle annular dark-field scanning transmission electron microscopy (HAADF-STEM) image of the WTe2 layer in the device viewed along the *b*-axis confirming the reduced lateral mirror symmetry in the *ac*-plane. The single crystal WTe2 layer exhibits lattice spacings of 3.6Å and 7.3Å along the *a*-axis and *c*-axis, respectively, with no observable stacking faults.

Next, we move to the discussion of wide-field quantum sensing measurements. Figure 1d illustrates the atomic structure of $V_B^-$ centers formed in hBN hexagonal crystal lattice with alternating boron (blue) and nitrogen (green) atoms. Each boron atom vacancy $V_B^-$ (red) is surrounded by three nearest-neighboring nitrogen atoms and its energy level structure is shown in Fig. 1e. The negatively charged $V_B^-$ defect has an out-of-plane oriented $S = 1$ electron spin featuring the characteristic Zeeman effect induced "three-level" spin system, which can be optically addressed by measuring the spin-dependent photoluminescence (PL)[17,19,28-30]. By probing the energy splitting between PL peaks corresponding to the $m_s = +1$ and $m_s = -1$ spin state(s), the magnitude of local static magnetic field parallel to the $V_B^-$ spin axis can be quantitatively obtained[14–16]. Figure 1f presents a series of optically detected magnetic resonance (ODMR) spectra of $V_B^-$ centers measured under different values of out-of-plane magnetic field $B_{ext}$. The Zeeman effect separates the $m_s = -1$ and $m_s = +1$ spin state(s) by an energy gap of $2\tilde{\gamma}B_{ext}$, where $\tilde{\gamma}$ denotes the gyromagnetic ratio of $V_B^-$ centers. To achieve nanoscale quantum imaging of Fe3GaTe2 nanoflake(s) in the current study, we utilize wide-field quantum magnetometry techniques to detect the fluorescence of $V_B^-$ ensembles across the field of view projected on a CMOS camera[14,15]. Figure 1g shows a magnetic stray field $B_s$ map of the prepared vdW stacking device (device A) measured at 260 K. One can see that the atomically thick (~6.3 nm) Fe3GaTe2 sample shows robust magnetization, in agreement with the magneto-transport studies[10,26]. The demonstrated hBN-based quantum imaging platform provides an attractive viewport to investigate microscopic spin behaviors in 2D vdW magnets.

We now present electrical transport measurement results showing field-free deterministic magnetic switching of the perpendicular Fe3GaTe2 magnetization. The spin source WTe2 was theoretically predicted and recently demonstrated as a suitable candidate to generate out-of-plane polarized spin currents due to its reduced lateral mirror symmetry[8–10,18]. It has a mirror symmetry with respect to the *bc*-plane but not the *ac*-plane as shown in Fig. 2a. When an electrical charge current is applied along the low-symmetry crystallographic axis (*a*-axis) of WTe2, the broken symmetry enables generation of spin currents with out-of-plane polarization in addition to the conventional spin Hall effect (Fig. 2b)[8–10,18]. When an in-plane charge current flows along the high-symmetry axis (*b*-axis), where the mirror symmetry is preserved in the *bc*-plane, the spin Hall effect dominates and directions of the electric field, spin currents, and spin polarization are mutually orthogonal as shown in Fig. 2c[8,18]. In the current work, the specific crystal symmetry of WTe2 was evaluated by rotational anisotropy second harmonic generation (RA-SHG) measurements that has been shown to be powerful in examining symmetries in crystalline solids[31–

[34]. A typical RA-SHG polar plot of few-layer WTe$_2$ is presented in Fig. 2d, from which we can identify the *bc* mirror plane with zero SHG intensity, and hence, the crystal *a*-axis that is normal to this mirror plane[35].

Figure 2e presents the current driven anomalous Hall loop of the Fe$_3$GaTe$_2$ sample without the assistance of an external magnetic field (see Supplementary Information Note 3 for details). The electric current is applied along the low-symmetry axis of WTe$_2$, and the measurement temperature is 200 K. We observed robust field-free deterministic magnetic switching, where the measured Hall voltage signals of Fe$_3$GaTe$_2$ show positive/negative jumps at the critical write current pulses and the terminal magnetic state is determined by the current polarity. The field-free bipolar switching of the perpendicular vdW magnet is attributed to out-of-plane spins injected from WTe$_2$, which effectively compensate for the local magnetic damping and overcome the anisotropy of proximal Fe$_3$GaTe$_2$ magnetization. To confirm the physical picture discussed above, we further measured the shift of anomalous Hall loops of Fe$_3$GaTe$_2$ to characterize the effective field produced by the unconventional SOTs from WTe$_2$. One can see that the measured anomalous Hall loops exhibit a clear shift towards the negative (positive) direction when large positive (negative) electric current pulses ($I_p = \pm 6.5$ mA) are applied along the *a*-axis of WTe$_2$ as shown in Fig. 2f. Note that the presented anomalous Hall loops become significantly tilted and deviate from the original square shape with a reduced magnetic coercivity due to enhanced Joule heating under higher current applications. The amount of the field shift becomes larger with increasing amplitude of electrical pulse currents and the efficiency of the equivalent out-of-plane magnetic field is characterized to be ~$8.1 \times 10^{-6}$ G/(A·cm$^{-2}$), from which the unconventional SOT efficiency is estimated to be $0.070 \pm 0.005$ in the all-vdW material system (see Supplementary Information Note 4 for details).

In sharp contrast, the signature of field-free deterministic magnetic switching vanishes when electric current pulses are applied along the high-symmetry crystallographic axis (*b*-axis) of WTe$_2$ as shown in Fig. 2g. Without the assistance of out-of-plane spins or an auxiliary in-plane magnetic field breaking the time reversal symmetry, the Fe$_3$GaTe$_2$ magnetization is driven to an intermediate phase that is in-plane magnetized or thermally randomized under both large positive and negative current pulses (see Supplementary Information Note 3 for details). The in-plane antidamping SOT could tilt the Fe$_3$GaTe$_2$ magnetization to the sample plane[8,36] followed by spontaneous out-of-plane self-remagnetization after electrical write current pulses are turned off. Theoretically, the incipient Fe$_3$GaTe$_2$ domains equally favor the two magnetic easy states, magnetization pointing up and pointing down, leading to fully randomly oriented magnetic domains without showing a total net perpendicular magnetization[8]. Current-induced Joule heating effect may also be involved in this process by driving thermal demagnetization in the Fe$_3$GaTe$_2$ sample, however, the overall "two-step" out-of-plane demagnetization-remagnetization physical picture remains valid due to the lack of a symmetry breaking field to deliver the deterministic perpendicular switching[8,36]. Note that no notable shift of anomalous Hall loops of the Fe$_3$GaTe$_2$ sample is observed when large electric current pulses are applied along the high-symmetry crystallographic axis of WTe$_2$ as shown in Fig. 2h, corroborating the absence of out-of-plane polarized spin currents.

We now utilize hBN-based wide-field quantum microscopy[14,15] to visualize the field-free deterministic magnetic switching of Fe$_3$GaTe$_2$ at the nanoscale. Figures 3a-3h present a series of representative stray field maps of the Fe$_3$GaTe$_2$ sample measured at the corresponding points ("A" to "H") on the unconventional SOT-driven magnetic hysteresis loop. Electric current pulses $I_p$ are applied along the low-symmetry axis of WTe$_2$ to ensure the field-free nature of the observed magnetic switching behavior (Fig. 3i). One can see that the $Fe_3GaTe_2$ sample starts from one of the spontaneous magnetic easy states with magnetization pointing down, producing a negative stray field in point "A" (see Supplementary Information Notes 5 and 6 for details). When electrical write current pulse $I_p$ is below the threshold value, the measured stray field $B_s$ map basically remains the same, indicating a negligible effect of out-of-plane spins on the $Fe_3GaTe_2$ magnetization. When ramping $I_p$ above the critical value, the out-of-plane antidamping SOT starts to locally flip the perpendicular $Fe_3GaTe_2$ magnetization, generating stray field with opposite polarity. The observed field-free magnetic switching starts from incipient magnetic domains at locations where the energy barrier is the lowest and then extends to neighboring $Fe_3GaTe_2$ sample area as shown in Figs. 3b-3d. When reaching the top plateau value of the anomalous Hall loop, ~70% portion of the $Fe_3GaTe_2$ magnetic domain has been flipped (Fig. 3d) in absence of an external magnetic field, in agreement with our magneto-transport measurement results. Inverting the electrical write current pulses into the negative regime leads to a reversal of the magnetic switching polarity accompanied by retraction of the flipped $Fe_3GaTe_2$ magnetic domains (Figs. 3e-3g). Sweeping $I_p$ back to the starting point ("A") completes the entire anomalous Hall loop, and the measured $Fe_3GaTe_2$ stray field map is almost identical with that measured in the initial magnetic state. The presented hBN wide-field quantum imaging measurements directly visualize the out-of-plane spin driven field-free deterministic magnetic switching of an atomically thick perpendicular vdW magnet, revealing SOT-induced nanoscale domain wall nucleation and propagation on 2D flatland. Note that we also visualized field-free full magnetic switching in other $Fe_3GaTe_2$ samples, and the observed switching ratio is sensitive to variations of magnetic inhomogeneities/defects and dimensions of prepared vdW devices (see Supplementary Information Notes 7 and 8 for details).

Achieving deterministic and nonvolatile magnetization control constitutes a key challenge for developing modern spintronic memory and computation technologies[1,2]. Next, we utilize hBN-based wide-field quantum magnetometry techniques to evaluate the (in)deterministic nature of presented field-free control of $Fe_3GaTe_2$ magnetization. Figure 4a shows the "step-like" variations of anomalous Hall signals of the $Fe_3GaTe_2$ sample in response to a train of positive and negative current pulses, $I_p = \pm 6.5$ mA, applied along the $a$-axis of $WTe_2$. Robust deterministic magnetization switching feature is highlighted in a series of wide-field quantum images recorded after individual current pulse applications as presented Fig. 4b. One can see that the positive and negative large current pulses reliably switch the $Fe_3GaTe_2$ magnetic domains between two deterministic states, from magnetization up to magnetization down or vice versa, in agreement with the "step-like" variations of anomalous Hall signals shown in the bottom panel of Fig. 4a. When $I_p$ is increased to $\pm 8.5$ mA, the magneto-transport results basically remain the same as shown in Fig. 4c. However, a microscopic view provided by our quantum imaging measurements indicates that the observed magnetic switching of $Fe_3GaTe_2$ has already deviated from the perfect deterministic nature. A small portion of the $Fe_3GaTe_2$ magnetic domains cannot be switched reproducibly between the two magnetic easy states. Although field-free deterministic magnetic switching still dictates most of the $Fe_3GaTe_2$ magnetization, it is evident that some local magnetic patterns show random variations after every electrical writing cycle [Fig. 4(d)]. The observed experimental feature is attributable to current induced Joule heating effect, which introduces random demagnetization-remagnetization behavior at certain $Fe_3GaTe_2$ sample areas where the local transient temperature is driven above the Curie point under large current pulse applications.

The deterministic magnetization control disappears when $I_p$ is applied along the $b$-axis of $WTe_2$. Figure 4f presents the evolution of $Fe_3GaTe_2$ magnetic stray field maps in response to the positive and negative pulse current train ($I_p = \pm6.5$ mA) applied along the high-symmetry *b*-axis of WTe$_2$. In this case, fully randomly oriented magnetic domains are spontaneously formed over the entire Fe$_3$GaTe$_2$ sample area with a net out-of-plane magnetization approximately to zero due to the absence of out-of-plane spins injected from WTe$_2$. While the measured anomalous Hall signals basically fluctuate around the zero value with little variations (Fig. 4e, bottom), hBN quantum imaging reveals that the microscopic Fe$_3$GaTe$_2$ magnetic patterns have dramatically changed at the nanoscale during the indeterministic magnetization control process (see Supplementary Information Note 7 for details).

In summary, we have demonstrated hBN-based wide-field quantum imaging of field-free deterministic magnetic switching in a room-temperature 2D magnet. Current-induced nanoscale nucleation and evolution of magnetic domains in atomically thick Fe$_3$GaTe$_2$ are visualized by $V_B^-$ spin ensembles contained in the hBN encapsulation layer. We further reveal how the observed magnetic switching evolves from deterministic to indeterministic behavior due to the interplay between out-of-plane spins, in-plane spins, and Joule heating effects. Our wide-field imaging results show that the Fe$_3$GaTe$_2$ sample reliably switches between two deterministic magnetic states when current pulses are applied along the low-symmetry crystallographic axis of the spin source material WTe$_2$. The deterministic nature of the field-free magnetic switching starts to degenerate when the magnitude of the electrical write current is above a certain threshold value due to involvement of local thermal heating effect. The deterministic magnetic switching feature is clearly absent when electric current pulses are applied along the high-symmetry crystallographic axis of WTe$_2$. Our study sheds light on the microscopic details of unconventional SOT-driven magnetization dynamics, providing insights into future design, integration, and engineering of 2D spintronic devices. We expect that the recently discovered room-temperature vdW magnet Fe$_3$GaTe$_2$ will serve as a promising building block for this purpose[10,25–27]. The presented hBN-based vdW quantum sensing platform further opens new opportunities for exploring nanoscale spin transport and dynamic behaviors in a broad family of miniaturized 2D heterostructure systems[3,19,23,24].

**Data availability**. All data supporting the findings of this study are available from the corresponding author(s) on reasonable request.

**Acknowledgements**. This work was primarily supported by the U.S. Department of Energy (DOE), Office of Science, Basic Energy Sciences (BES), under award No. DE-SC0024870. J. Z. acknowledges the support from the U.S. National Science Foundation under award No. DMR-2342569. H. L. was supported by the Air Force Office of Scientific Research (AFOSR) under award No. FA9550-21-1-0125. C. R. D. also acknowledges the support from the Office of Naval Research (ONR) under grant No. N00014-23-1-2146. The growth and characterization of Fe$_3$GaTe$_2$ single crystals were supported by the Center on Programmable Quantum Materials, an Energy Frontier Research Center funded by the U.S. Department of Energy (DOE), Office of Science, Basic Energy Sciences (BES), under award No. DE-SC0019443. K. D. and R. C. were supported by the U.S. National Science Foundation under Award No. DMR-2339315. L. Z. acknowledges the support from the U.S. Department of Energy (DOE), Office of Science, Basic Energy Science (BES), under award No. DE-SC0024145. R. H. and N. A. acknowledge support from the National Science Foundation through the Materials Research Science and Engineering Center at the University of Michigan, Award No. DMR-2309029. The synthesis of WTe$_2$ crystals was supported by the UC Lab Fees Research Program (grant No. LFR-20-653926). The synthesis of $^{10}$B-enriched monoisotopic hBN crystals was supported by NASA-CLEVER SSERVI (CAN no. 80NSSC23M0229) and NASA-MSFC (CAN no. 80NSSC21M0271).

**Author contributions**. X. Z. performed the quantum sensing measurements and analyzed the data with J. Z., S. L., and H. L. J. Z. prepared the vdW stacking devices and performed electrical transport measurements. C. H. grew and characterized bulk Fe$_3$GaTe$_2$ crystals under the supervision of X. X. and J.-H. C. K. D. and R. C. contributed to theoretical input and discussions. C. W. performed the second harmonic generation measurements under the supervision of L. Z. N. A. performed the TEM characterizations under the supervision of R. H. H. J., S. X., and V. T. provided WTe$_2$ crystals. F. A. A., R. S. T., S. R., and Z. J. provided monoisotopic hBN crystals. C. R. D. and H. L. W. supervised this project.

**Competing interests**
The authors declare no competing interests.

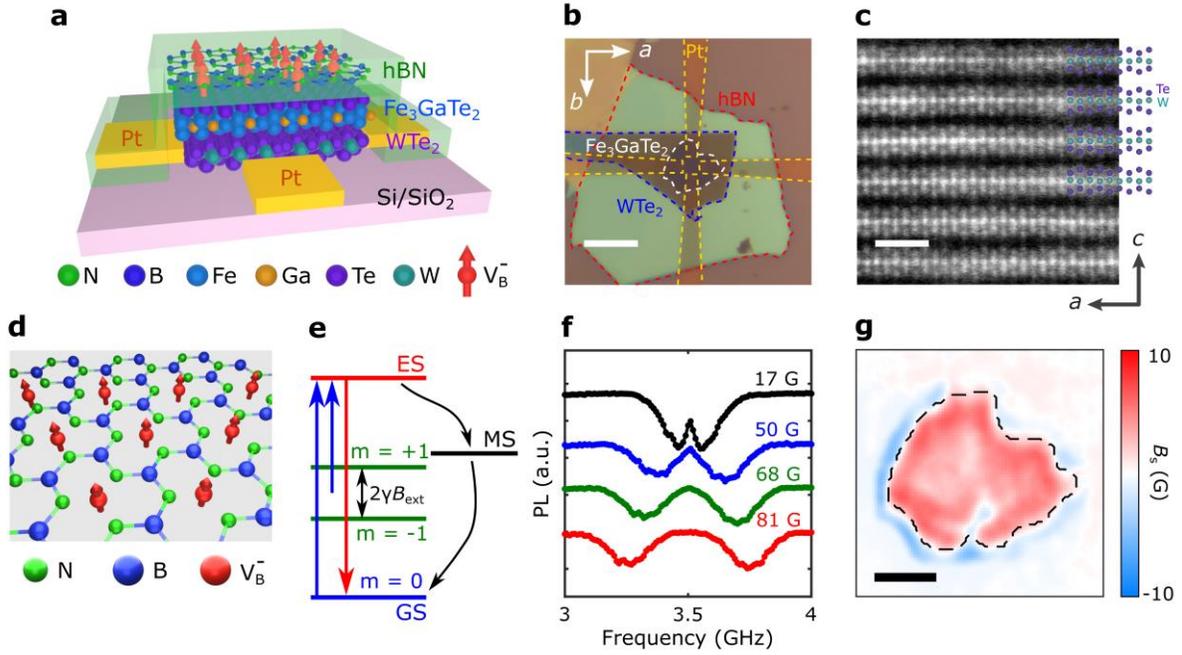

**Figure 1. Device structure and hBN-based wide-field quantum imaging platform. a**. Schematic of the WTe$_2$/Fe$_3$GaTe$_2$/hBN vdW heterostructure for quantum imaging and electrical SOT switching measurements. **b** Optical microscopy image of a prepared WTe$_2$/Fe$_3$GaTe$_2$/hBN device. The boundaries of constituent vdW layers are outlined by red (hBN), blue (WTe$_2$), and white (Fe$_3$GaTe$_2$) dashed lines. Pt electrodes are highlighted by yellow dashed lines. The scale bar is 10 μm. **c** Cross-sectional HAADF-STEM image of a prepared device viewed along the *b*-axis of WTe$_2$. The highlighted atomic structure shows the broken lateral mirror symmetry in the *ac*-plane of WTe$_2$. The scale bar is 1 nm. **d** $V_B^-$ spin defects (red arrows) formed in the hexagonal crystal structure with alternating boron (blue) and nitrogen (green) atoms. **e** Energy level diagram of a $V_B^-$ spin defect and optical excitation (blue arrow), radiative recombination (red arrow), and nonradiative decay (black arrow) between the ground state (GS), excited state (ES), and metastable state (MS). **f** A series of $V_B^-$ ODMR spectra measured at different values of a perpendicular magnetic field. **g** A stray field map of the Fe$_3$GaTe$_2$ sample measured at 260 K with an external perpendicular magnetic field $B_{ext}$ of 30 G. Black dashed lines outline the boundary of the Fe$_3$GaTe$_2$ flake, consistent with the device details shown in Fig. 1b. The scale bar is 4 μm.

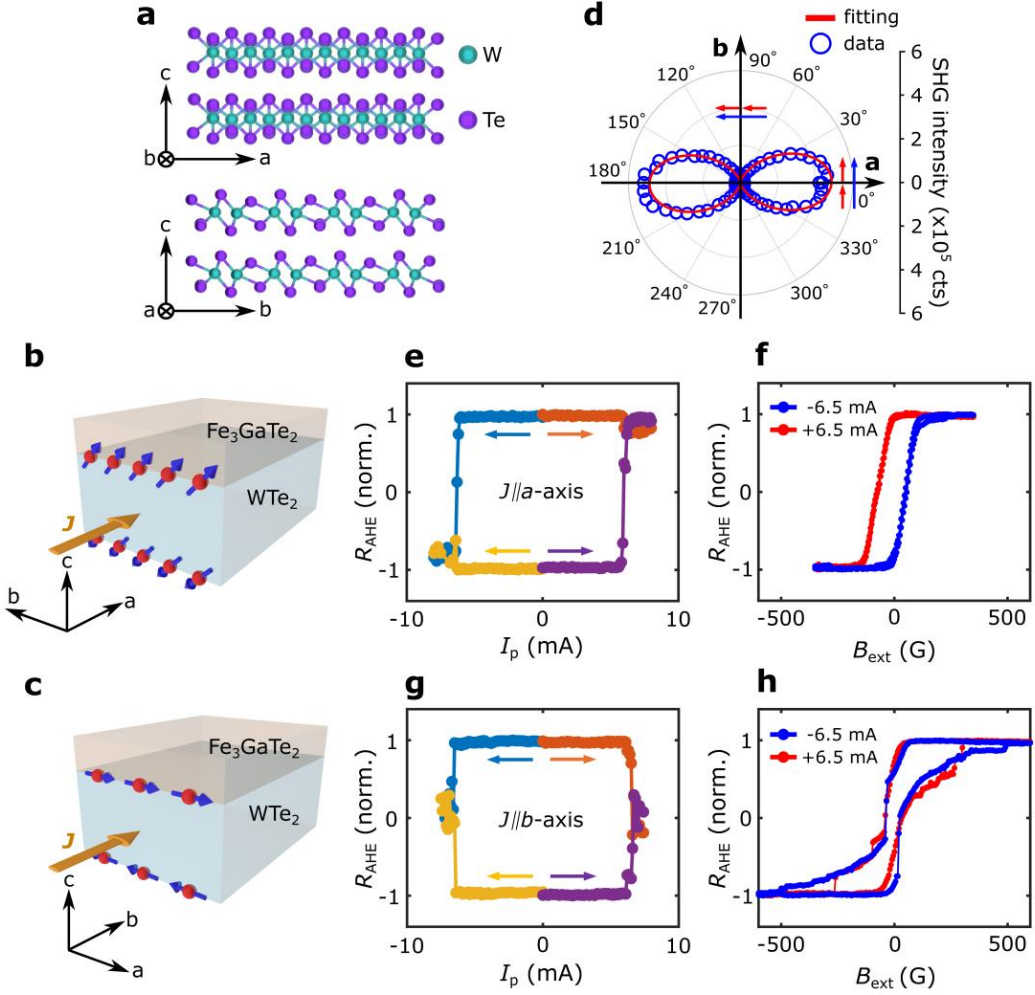

**Figure 2. Field-free deterministic and indeterministic magnetic switching in WTe₂/Fe₃GaTe₂/hBN. a** Atomic arrangement of WTe$_2$ viewed along its *b*-axis (top) and *a*-axis (bottom) directions. **b, c** Generation of out-of-plane and in-plane polarized spin currents when an electrical charge current *J* flows along the low-symmetry and high-symmetry crystallographic axes of WTe$_2$. **d** An azimuthal angle dependent SHG pattern in parallel channel characterizing the reduced crystal symmetry in the *ac*-plane of WTe$_2$. The fitting is based on the electric-dipole SHG response under the *m* point group. Red and blue arrows indicate the polarizations of incident and reflected light at the 0° and 90°, respectively. **e, g** Normalized anomalous Hall resistance of Fe$_3$GaTe$_2$ measured as a function of electrical write current pulse $I_p$ applied along the *a*-axis (**e**) and *b*-axis (**g**) of WTe$_2$. The magnitude of $I_p$ was increased from zero with the sweeping direction indicated by the color arrows. Fe$_3$GaTe$_2$ was initialized to the $m_z = 1$ or $m_z = -1$ state in the beginning of each sweep sequence. External auxiliary magnetic fields were absent in the presented SOT measurements. **f, h** Variations of measured Fe$_3$GaTe$_2$ anomalous Hall loops in response to positive and negative electric current pulses ($I_p = 6.5$ mA) applied along the *a*-axis (**f**) and *b*-axis (**h**) of WTe$_2$. The presented magneto-transport measurements were performed at 200 K.

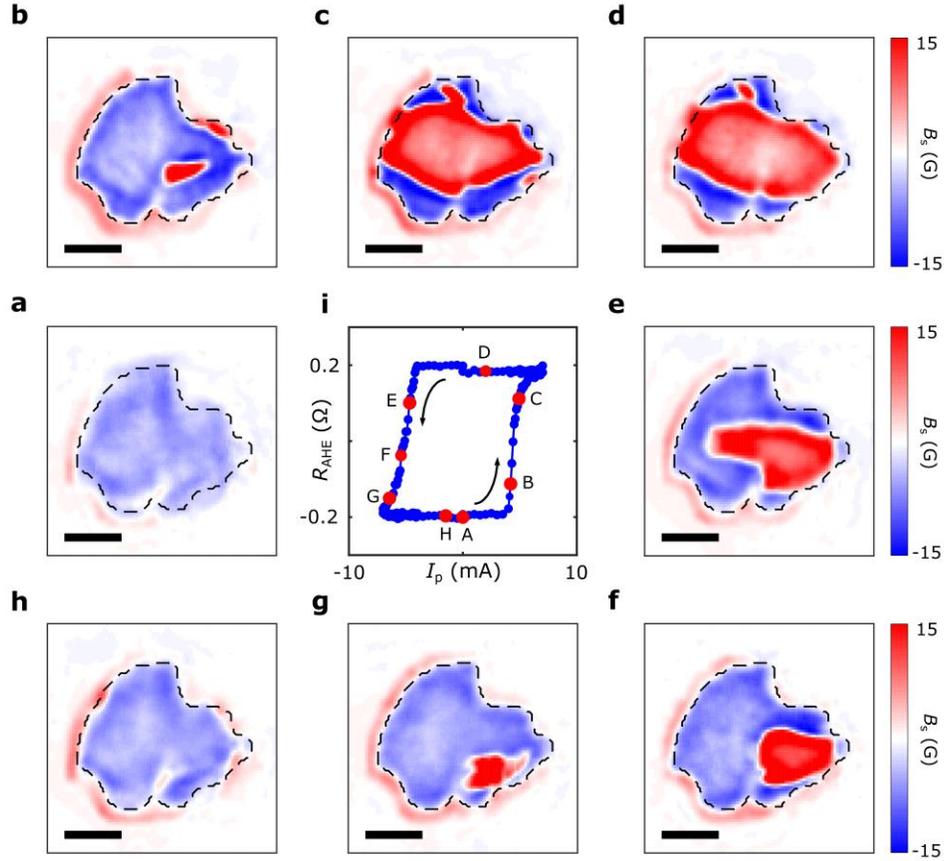

**Figure 3. Visualizing field-free deterministic magnetic switching of $Fe_3GaTe_2$. a-h** hBN-based quantum imaging of microscopic evolutions of $Fe_3GaTe_2$ magnetic domains during the field-free deterministic magnetic switching process. The scale bar is 4 μm. **i** Anomalous Hall resistance of $Fe_3GaTe_2$ measured as a function of electrical write current pulse $I_p$ in absence of an external magnetic field. The arrows indicate that $I_p$ was swept from zero following the counterclockwise direction around the hysteresis loop and finally returned to the starting point. hBN-based quantum imaging measurements presented in Figs. 3a-3h were performed at the corresponding points from "A" to "H" marked on the field-free deterministic switching loop (Fig. 3i). The SOT-driven magnetic switching and quantum sensing measurements were performed at 260 K.

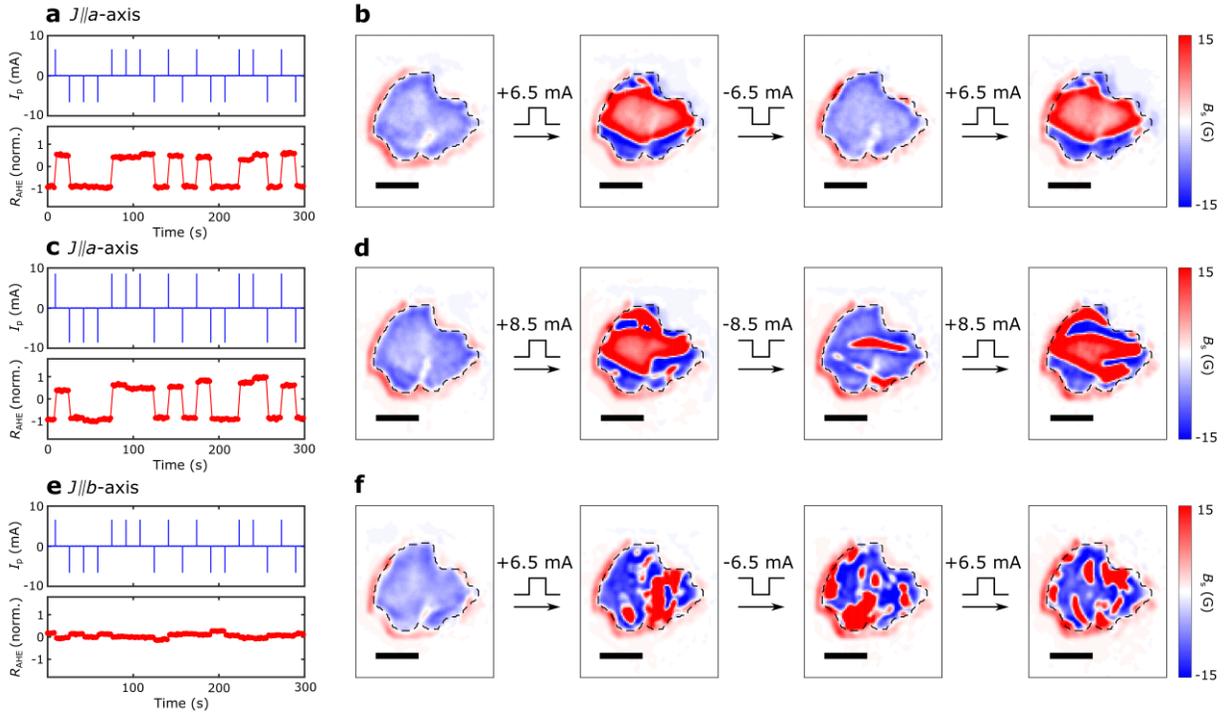

**Figure 4. Evaluating (in)deterministic nature of field-free magnetization control of $Fe_3GaTe_2$.**
**a, c** Deterministic magnetic switching of $Fe_3GaTe_2$ using a train of positive and negative electric current pulses ($I_p$) with a magnitude of 6.5 mA (**a**) and 8.5 mA (**c**) applied along the *a*-axis of $WTe_2$. **b, d** Variations of magnetic stray field patterns of the $Fe_3GaTe_2$ sample after individual electric current pulse applications along the *a*-axis of $WTe_2$ for $I_p = \pm6.5$ mA (**b**) and $\pm8.5$ mA (**d**). **e, f** Electrical SOT measurements and the corresponding quantum imaging results for a series of current pulses ($I_p$) with a magnitude of $\pm6.5$ mA applied along the *b*-axis of $WTe_2$. Scale bar is 5 μm in all figures. The SOT-driven magnetic switching and quantum sensing measurements were performed at 260 K.

13